

\def\tb{\bar \theta}
\def\pbar{\Gamma^\mu P_\mu}
\def\ul{\underline}
\def\np{\vfill\eject}
\def\nl{\hfill\break}
\def\ft#1#2{{\textstyle{{#1}\over{#2}}}}
\def\narrow{\advance\leftskip by 40pt \advance\rightskip by 40pt}
\def\nonarrower{\advance\leftskip by -40pt\advance\rightskip by -40pt}
\def\AB{\bigskip \centerline{\bf ABSTRACT}\medskip\narrow}
\def\AE{\bigskip\nonarrower}
\def\singlespace{\normalbaselines}

\font\titlefont=cmbx10 scaled\magstep2
\font\authorfont=cmr12
\font\fontina=cmr9
\font\bigi=cmbx10 scaled\magstep2
\baselineskip 15pt
\magnification=\magstep1

\rightline{CTP TAMU--58/94}
\rightline{November 1994}
\rightline{ hep-th@xxx/9411055}

\vskip 2truecm
\centerline{\titlefont Spacetime and Worldvolume Supersymmetric}
\centerline{{\titlefont Super}\ \ \bigi p--{\titlefont Brane Actions}
 \footnote{$^*$}{\fontina \sl Based on a talk presented at the
  VIth Regional Conference in Mathematical Physics, \nl
\phantom\quad\quad Pakistan, 5-11 February 1994. }}
\vskip 1truecm
\centerline{{\authorfont E. Sezgin}\footnote{$^\dagger$}{\fontina
\sl  Supported in part by the National Science Foundation, under grant
PHY-9106593.}}\vskip 1truecm
\noindent{\sl  Department of Physics,
Texas A\&M University, College Station, TX 77843--4242, USA.}
\vskip 1.5truecm
\AB\singlespace

We review the salient features of spacetime {\it and} worldvolume
supersymmetric super $p$--brane actions. These are
sigma models for maps from a worldvolume superspace to the target
superspace. For $p$--branes, the symmetries of the model depend
crucially on the existence of {\it closed super $(p+1)$--forms on a
worldvolume
superspace}, built out of the pull-backs of the Kalb-Ramond super $(p+1)$--form
in target superspace and its curvature. This formulation of super
$p$--branes is usually referred to as the twistor-like formulation.

\AE

\np

\noindent{\bf 1. Introduction}

\bigskip

Manifest spacetime supersymmetry is clearly a desirable feature to have in a
superstring theory. Although the Green-Schwarz formulation of superstring goes
a long way in this direction [1], nobody really knows how to quantize the
theory. This problem is not so much due to the fact that we are dealing with
string theory. Even the manifestly supersymmetric formulation of a
superparticle
suffers from this problem. Needless to mention, the manifestly supersymmetric
formulation of higher super $p$--branes [2-4] do not fair any better either.

The difficulty in covariant quantization of super $p$--branes is essentially
due to the fact that the so called $\kappa$--symmetry [1] of these theories is
an infinitely reducible one. However, this symmetry is needed in
order to arrive at non-manifest worldvolume supersymmetry after gauge
fixing. It is natural to enquire into the possibility of dispensing
with $\kappa$--symmetry and building a new kind of action which will have
manifest target space {\it and} worldvolume supersymmetry from the
beginning. In such a formulation, one may hope to get rid of the infinite
reducibility problem and thus  making a progress towards covariant
quantization.

There is another outstanding problem in super $p$--branes
which might also be more easily solvable in these new type of
formulations, namely, the problem of how to couple Yang-Mills fields to them.
In particular, the case of super fivebrane is of great interest, because there
is some evidence for its being closely related to the heterotic superstring in
ten dimensions [5].

The motivations mentioned above, so far have only led to new and
interesting formulations of super $p$--branes. Unfortunately, we still don't
know if the covariant quantization and Yang-Mills coupling problems have
solutions in these new formulations. In fact, in the general case, we don't
even
know what the true physical degrees of freedom are in a physical gauge.
Clearly, a lot of work remains to be done. The purpose of this note is to
briefly survey what alternative formulations exist, and then focusing  on the
twistor-like formulation, giving the main results. We shall begin with the
massless superparticle (known as the Brink-Schwarz superparticle [6]), and
then, after discussing its various twistor formulations [7-12], we move on to
massive superparticles which resemble higher super $p$--branes in many
respects. We shall conclude with the summary of results for super
$p$--branes.

\bigskip

\noindent{\bf 2. Brink-Schwarz Superparticle}

\bigskip

To illustrate the main ideas involved in manifestly supersymmetric formulation
of strings and higher extended objects, it is useful to consider the simplest
case of a massless superparticle in ten dimensions. The relevant action is due
to Brink and Schwarz [6], and is given by
$$
S=\int d\tau \left[ P_\mu({\dot X}^\mu-i{\bar \theta}\Gamma^\mu{\dot
\theta})-{1\over 2}e P_\mu P^\mu\right]\ ,   \eqno(2.1)
$$
where $e(\tau)$ is the einbein on the worldline, $(X^\mu, \theta^\alpha)
(\mu=0,1,...,9; \alpha=1,...,16)$ are the coordinates of target superspace, and
${\dot \theta}\equiv d\theta/d\tau$. This action has  manifest spacetime {\it
global} supersymmetry, and a nonmanifest {\it local} worldline symmetry known
as the $\kappa$-- symmetry given by [1]
$$
\eqalign{
\delta X^\mu=& i\tb \Gamma^\mu \delta \theta\ ,   \cr
 \delta\theta=& \Gamma^\mu P_\mu \kappa\ , \cr
\delta e=& 4i{\bar \kappa}{\dot \theta}\ , \cr
\delta P_\mu=& 0\ . \cr} \eqno(2.2)
$$
The trouble with this action is that we don't know how to covariantly quantize
it. The easiest way to see the problem is to note that the transformation rule
for the fermionic variable $\theta$ has the zero mode $\kappa_0=\pbar
\kappa^1$ modulo the zero mode $\kappa^1_0=\pbar \kappa^2$ modulo the zero
mode $\kappa^2_0=\pbar \kappa^3$, ad infinitum. This shows
that we are dealing with an infinitely reducible gauge symmetry, requring the
introduction of infinitely many ghost fields. There are ambiguities in
calculations involving infinitely many ghost fields, and moreover,
additional complications arise due to the fact that the
residual gauge symmetries are not completely fixed by the covariant
quantization procedure.

Attempts have been made to introduce new variables which would make it possible
to reformulate the superparticle action so that, while the physical degrees of
freedom are the same, the $\kappa$ symmetry, and hence the attendant
quantization problems are removed. A number of such reformulations invariably
involve the so called twistor or twistor-like variables. These variables are
closely related to certain representations of appropriate superconformal
groups.
Ultimately, we will be interested in one such formulation [10] which is the
most
suitable one for generalization to higher super $p$--branes. However, in order
to stress the differences between various twistor formulations, we
shall briefly review three such formulations below [7,9,10].
\bigskip
\noindent{\bf 3. Three Different Twistor Formulations of the Massless
Superparticle}
\bigskip
\noindent{\it (3a)  Supertwistor Formulation }
\bigskip
This formulation works in dimensions where a superconformal
group exists [7]. Hence the restriction to $d\le 6$. Let us consider, for
example, the case of $N=8$ supersymmetric particle in $d=3$. The main idea of
the supertwistor formulation is to introduce a supertwistor variable, which is
simply a finite dimensional representation of of the superconformal group, in
this case $OSp(8|4)$, and consists of $(\omega_\alpha, \lambda_\alpha,
\psi^i)$, where $\omega_\alpha$ and $\lambda_\alpha$ are {\it commuting}
$SL(2,R)$ spinors, while $\psi^i$ is an anticommuting $SO(8)$ vector. These
variables allow us to perform the following field redefinitions:
$$
\eqalign{
P^\mu=&\Gamma^\mu_{\alpha\beta}\lambda^\alpha \lambda^\beta\ , \cr
\psi^i=&\theta^{\alpha i} \lambda_\alpha\ , \cr
\omega^\alpha=&X^\mu\Gamma_\mu^{\alpha\beta} \lambda_\beta +i\theta^{\alpha i}
\psi_i\ . \cr} \eqno(3.1)
$$
It should be noted that the variable $P^\mu$ as defined above, satisfies the
constraint $P^\mu P_\mu=0$. In terms of the supertwistor variables, one can
write down the action
$$
S=\int d\tau ({\dot \omega}^\alpha \lambda_\alpha + i{\dot \psi}^i
\psi^i)\ . \eqno(3.2)
$$
We no longer need to consider the $\kappa$ symmetry of this action,
because, as one can easily check, the supertwistor variables themselves
are $\kappa$ invariant. Furthermore one can check that this action produces the
first  two terms of the Brink-Schwarz action (2.1), while the consequence
of the last term, namely $P^\mu P_\mu=0$ is automatically satisfied
due to the superstwistor form of the variable $P^\mu$.

The above action has a very simple form, and it can actually be covariantly
quantized [7]. However, the cases of $d=4$ and $d=6$ are more complicated,
because new type of local bosonic symmetries emerge [7,8]. Moreover, the above
formalism doesn't generalize to $d=10$, since there is no superconformal group
in ten dimensions.
\bigskip
\noindent{\it (3b) Supertwistor-like Formulation}
\bigskip
Although a supreconformal group doesn't exist in ten dimensions, one may
nonetheless try to use variables similar to the supertwistor variables
described in the previous section. This has been done by  Berkovits [9] who
used the following variables
$$
\eqalign{
P^\mu=&\lambda^\alpha\Gamma^\mu_{\alpha\beta}\lambda^\beta\ ,\cr
\psi^\mu=&\lambda^\alpha\Gamma^\mu_{\alpha\beta}\theta^\beta\ ,\cr
\omega^\alpha=&
X^\mu\Gamma_\mu^{\alpha\beta}\lambda_\beta-
i\psi^\mu\Gamma_\mu^{\alpha\beta}\theta^\beta\ , \cr}\eqno(3.3)
$$
Note the similarity with the set of variables defined in (3.1). These
variables, however, don't corespond to a superconformal group, and therefore,
we shall refer to them as supertwistor-like variables. Unlike the
supertwistors, they are only partially $\kappa$-invariant. In terms of these
variables, Berkovits' action reads
$$
\eqalign{
S=&\int d\tau \big[ -2\omega^\alpha {\dot \lambda}_\alpha
-2i\psi^\mu{\dot \psi}_\mu + h\psi^\mu{\bar \lambda}\Gamma_\mu \lambda \cr
& +h_\alpha \left({\bar \lambda}\Gamma^\mu \lambda
\Gamma_\mu^{\alpha\beta}\omega_\beta-2{\bar \lambda}\omega \lambda^\alpha
+2i\psi^\mu \psi^\nu \Gamma_{\mu\nu}^{\alpha\beta}
\lambda_\beta\right)\big]\ , \cr} \eqno(3.4)
$$
where $h$ and $h_\alpha$ are Lagrange multiplier fields. While the
covariant quantization of this model is possible [9], its generalization to
string theory is not so obvious. It turns out that one way to achieve this is
to  do away with $\omega_\alpha$ and $\psi^\mu$ type variables, but
keeping the variable $\lambda_\alpha$ which continues to play a central role
in passing to a formulation where $\kappa$ symmetry is traded for worldline
supersymmetry. This is achieved in the twistor-like formulation described
below [9].
\bigskip
\noindent{\it (3c)\  Twistor-like Formulation}
\bigskip
This formulation is due to Sorokin, Tkach, Volkov and Zheltuhkin [10], and it
maintains only the twistor-like variable $\lambda$, as defined in (3.3). The
advantage of doing so will become clear below. For simplicity, let us focus
on the case of $d=3, N=1$ massless superparticle for now. The Brink-Schwarz
action is  replaced by
$$
\int d\tau P_\mu\left( {\dot X}^\mu-i\tb \Gamma^\mu {\dot \theta}+{\bar
\lambda}\Gamma^\mu\lambda \right)\ . \eqno(3.5)
$$
{}From the $\lambda$ equation of motion,
$P^\mu\Gamma^\mu\lambda=0$, one finds the solution $P_\mu={\bar
\lambda}\Gamma_\mu\lambda$,  which satisfies $P^\mu P_\mu=0$, thanks to the
identity: $\Gamma_{\mu(\alpha\beta}\Gamma^\mu_{\gamma\delta)}=0$.
This action has {\it worldline\ local\ n=1\ supersymmetry} which replaces the
$\kappa$ symmetry. The $n=1$ supersymmetry closes on-shell. To close it
off-shell, one introduces superfields on the world superline as follows [10]:
$$
\eqalign{
P_\mu(\tau,\eta)=&P_\mu(\tau)+i\eta Q_\mu(\tau)\ , \cr
X^\mu(\tau,\eta)=& X^\mu(\tau)+iY^\mu(\tau)\ , \cr}\eqno(3.6)
$$
where $(Q_\mu, Y^\mu)$ are auxiliary fields and $\eta$ is the fermionic
coordinate in the $n=1$  worldine superspace. We can view the twistor variable
$\lambda$ as the superpartner of the target superspace fermionic coordinate
$\theta$ and define
$$
\theta(\tau,\eta)=\theta(\tau)+\eta \lambda(\tau)\ . \eqno(3.7)
$$
Then, the off-shell version of the action (3.5) can be written as a superspace
integral [10]
$$
S\int=-i d\tau d\eta P_\mu\left( DX^\mu+i{\bar \theta}\Gamma^\mu D\theta
\right)
\ , \eqno(3.8)
$$
where $D={\partial\over \partial\eta}+i\eta{\partial\over \partial\tau}$. One
can furthermore combine $X^\mu(\tau,\eta)$ and $\theta(\tau,\eta)$ to define
target superspace coordinates, which are worldline superfields.

Note that the action (3.5) is linear in time derivative and doesn't contain
the einbein. Thus, it has the form of a Wess-Zumino term. Moreover, it turns
out
that this form of the action does admit generalization to higher $n(N)$
supersymmetry, curved superspace as well as superstrings and higher super
$p$--branes. First, let us describe the higher $n(N)$ supersymmetry.

A convenient notation for dealing with two superspaces is as follows. We
use the same letters for worldline and target superspaces, but
distinguish the two by underlying the target superspace coordinates.  The
notation for general super $p$--branes can be summarized as follows:
\bigskip
\settabs 6 \columns
\+  &Worldline superspace:&&${\cal M}$\ : &$Z^M= (X^m, \theta^\mu)$ & \cr
\+ &Target superspace:&&${\ul {\cal M}}$\ : &$Z^{\ul M}=(X^{\ul m},\theta^{\ul
\mu})$&\cr \+ &Worldline supervielbein:&&$E_M{}^A$\ : & $A=(a,\alpha)$ \cr
\+ &Worldline supervielbein:&&$E_{\ul M}{}^{\ul A}$\ : &
      ${\ul A}=({\ul a},{\ul \alpha})$\ . \cr
\bigskip
For the superparticle the
target superspace notation is as above, but for the worldline superspace we use
the notation, $Z^M=(\tau, \theta^\mu),\  A=(0,r),\ (\mu,r=1,...,n)$.
For super $p$--branes, on the other hand, the range of indices are as
follows: $m,a=0,1,...,p$;  $\mu,\alpha=1,...,n$; ${\ul m,\ul
a}=0,1,...,d-1$; ${\ul \mu,\ul \alpha}=1,...,MN$, where $M$ is the dimension of
the minimum dimensional spinor representation of $SO(d-1,1)$ and $N$ is the
dimension of the defining representation of the automorphism group ${\ul G}$ of
the super Poincar\'e algebra in $d$ dimensions. It is important to note that
the
number of worldvolume supersymmetries is half of the target space
supersymmetries (counted in terms of worldvolume spinors), i.e.
$n={1\over 2}MN$.  A further notation is that when the automorphism group is
nontrivial, the index $\alpha$ represents a pair of indices $\alpha'r$. Thus,
$\Gamma^a_{\alpha\beta}=\Gamma^a_{\alpha'\beta'}\eta_{rs}$, where $\eta_{rs}$
is the invariant tensor of the automorphism group $G$ of the worldvolume super
Poincar\'e algebra. For further aspects of this notation, see the Table.

Let us now consider the case of $n=8$ supersymmetric massless superparticle
[12,13].
The twistor-like variable $\lambda_r{}^{\ul \alpha}$ satisfies the constraint
$$
{\ul \lambda}_r\Gamma^{\ul a}\lambda_s={1\over 8}\delta_{rs}\left( {\bar
\lambda}_q\Gamma^{\ul a}\lambda_q\right)\ . \eqno(3.9)
$$
We use a notation in which the contracted ${\ul \alpha}$ indices are
supressed, and the parenthesis such as those in (3.9) indicate these
contractions. It is useful to express the constraint (3.9) in
geometrical way. To this end, let us define
$$
E_A^{~\ul A}=E_A^{~M}\big( \partial_M Z^{\ul M}\big) E_{\ul M}^{~\ul A}\ .
\eqno(3.10)
$$
We can make the identifications
$$
   E_r^{\ul{\alpha}}\vert_{\theta=0}=\lambda_r^{\ul{\alpha}}\ ,\quad\quad
  E_0^{\ul a}\vert_{\theta=0}={\cal E}_0^{\ul a}\ , \eqno(3.11)
$$
where ${\cal E}_0^{\ul a}={\dot X}^a-i{\bar \theta}\Gamma^a{\dot
\theta}$. Thus, we have the expansion $\theta^{\ul
\alpha}(\tau,\theta)=\theta^{\ul\alpha}(\tau)
+\lambda_r^{\ul\alpha}(\tau)\theta^r+\cdots $. The action can now be
written as follows
$$
S=\int d\tau d^8\theta P_{\ul a}{}^r E_r{}^{\ul a}\ . \eqno(3.12)
$$
The field equation for the Lagrange multiplier superfield $P_{\ul a}{}^r$ is
$$
E_r{}^{\ul a}=0\ . \eqno(3.13)
$$
Taking the supercurl of this equation, we arrive at the integrability condition
$$
      (E_r\Gamma^{\ul a} E_s) =\delta_{rs} E_0{}^{\ul a}\ .\eqno(3.14)
$$
the lowest component of
which implies the constraint (3.9). In deriving (3.14), the
following Lemma is useful:
$$
D_A E_B^{~\ul C}-(-1)^{AB}D_B E_A^{~\ul C}=-T_{AB}{}^C
E_C^{\ul C}+(-1)^{A(B+{\ul D})} E_B^{~\ul D}
E_A^{~\ul E}T_{{\ul E}{\ul D}}{}^{\ul C} \ , \eqno(3.15)
$$
where the covariant derivative $D_A=E_A^{~M}D_M$ rotates the indices $A$
and ${\ul A}$ and the tangent space components of the supertorsion
$T_{MN}{}^C=\partial_M E_N^{~C}+\Omega_M^{CD}
E_{ND}-(-1)^{MN}(M\leftrightarrow N)$ are defined as:
$
T_{AB}{}^C=(-1)^{A(B+N)}E_B^{~N}E_A^{~M}T_{MN}{}^C
$,
and similarly for $T_{{\ul A}{\ul B}}{}^{\ul C}$.

There may seem to
be many more fields in the $\theta$--expansion of the action functional in
(3.12). However, as shown in [12,13], there are many redundant fields, and the
action (3.12) is  classically equivalent to the Brink-Schwarz superparticle
action. For a detailed discussion of this equivalence as well as the local
supersymmetry of the action and its relation to $\kappa$-symmetry, we refer the
reader to [12,13].

	One immediate bonus that follows from the above formulation is that its
generalization to curved superspace is immediate. We simply elevate the
supervielbeins occuring in the action formula to those of curved superspace.
The action then has precisely the same form as in the flat superspace. As for
the twistor constraint (3.14), it will now follow from the integrability
condition of (3.13), followed by imposition of a suitable set of supertorsion
constraints (both in worldline and target superspaces). In what follows,
our strategy will be to fix the  geometry  of the worldvolume and target
superspaces from the beginning, though one may try to determine them, at least
partially, from other considerations such as worldvolume supersymmetry.

Another bonus of the twistor-like formulation is that it generalizes quite
naturally to superstrings [13,14] and higher super $p$--branes [15,16] . The
case of heterotic string has been discussed in great detail in Refs. [13,14].
Here, we shall  focus
on the results for higher super $p$-branes [15,16]. Since higher super
$p$--branes resemble in many respects the massive superparticle, we shall first
decsribe the latter case, and then give the result for the general super
$p$--branes. For an alternative and distinct approach to
twistor-like formulation of super $p$--branes, see Ref. [17].
\bigskip
\noindent{\bf 4. Twistor-like Formulation of the Massive Superparticle}
\bigskip
The usual $\kappa$ invariant massive superparticle action is given by
$$
    S=\int d\tau \bigg(\ft12 e^{-1}{\cal E}_\tau^{\ul a}
{\cal E}_\tau^{\ul a}  +\ft12  e + {\cal E}_\tau^{\ul A}B_{\ul A}\bigg)\ ,
\eqno(4.1)
$$
where $e$ is the einbein on the worldline, ${\cal E}_\tau{}^{\ul
A}=\partial_\tau Z^{\ul M} E_{\ul M}{}^{\ul A}$ and $B_{\ul A}=E_{\ul
A}{}^{\ul M} B_{\ul M}$. The latter is the super one-form that is analogous
to the Kalb-Ramond field in string theory. It is needed for the
$\kappa$--symmetry of the action. This symmetry imposes some constraints on
the supertorsion as well as $H=dB$. The form of these constraints and the
$\kappa$ symmetry transformation rules can be found in [18]. Here, we shall
make further choices regarding the form of the constraints, in order to fix
the target superspace geometry as much as possible. We will work with the
following constraints
$$
\eqalign{
T_{\ul{\alpha\beta}}{}^{\ul c}=&
-2i(\Gamma^{\ul c})_{\ul{\alpha\beta}}\ ,\cr
T_{\ul{\alpha b}}{}^{\ul c}=&0\ , \cr
H_{\ul{\alpha\beta}}=&-2i C_{\ul{\alpha\beta}}\ , \cr
H_{\ul {\alpha a}}=&0  \ .\cr}\eqno(4.2)
$$

For later purpose, it is also useful to give the Nambu-Goto form of the
action, which is obtained from (4.1) by eliminating the einbein through its
equation of motion
$$
S=\int d\tau\bigg[ ({\cal E}_\tau^{\ul a}{\cal E}_\tau^{\ul a})^{1/2}
+{\cal E}_\tau^{\ul A}B_{\ul A}\bigg]\ . \eqno(4.3)
$$

In order to pass to the twistor-like formulation, we adopt the target
superspace constraints (4.2), and elevate the worldline to an $N=8$
superspace with the following constraints
$$
\eqalign{
T_{rs}{}^0=&-2i\delta_{rs}\ ,\cr
T_{0r}{}^0=&0\ ,\cr
T_{s0}{}^r=&0\ ,\cr
T_{rs}{}^q=&0\ .  \cr} \eqno(4.4)
$$
The rationale behind these constraints is that they still leave room for
$N=8$ local superdiffeomorphisms [12,13]. Having now specified the geometry
of worldline and target superspaces, in addition to the superfields occuring
in the action (3.12) and the super one-form $B_M$, we introduce two Lagrange
multiplier superfields $P^M$ and $Q$, and propose the following action [16]
$$
S= \int d\tau d^8 \theta\bigg[ P_{\ul a}^r E_r^{\ul a}
+P^M({\tilde B}_M -\partial_M Q)\bigg]\ ,  \eqno(4.5)
$$
where the super one-form ${\tilde B}$ is defined as
$$
{\tilde B}_M= \partial_M Z^{\ul M}B_{\ul M} -{i\over 16}E_M^0 H_{rr}
\ , \eqno(4.6)
$$
and $H_{rr}=E_r^{\ul A}E_r^{\ul B} H_{\ul{BA}}$,
$H_{\ul{BA}}$ are the tangent space components of the field strength
$H=dB\ :
H_{\ul {AB}}= (-)^{\ul {A}(\ul {B} + \ul {N})} E_{\ul {B}}^{\ul {N}}
E_{\ul {A}}^{\ul {M}} H_{\ul {MN}}
$, where the indices in the exponent indicate Grassmannian parities. Recall
that $M=(\tau,\mu)$, $A=(0,r)$, ${\ul M}=(\ul m,\ul \mu)$ and ${\ul A}=(\ul
a,\ul \alpha)$. The indices of the bosonic (fermionic) coordinates have the
parity 0(1).

The form of the action (4.5) is
inspired by results results of Refs. [12,13] for massless superparticle
and heterotic superstring. Note that the
independent world-line {\it superfields} in the action are:
$P_{\ul a}^r,\ P^M,\
Q,\ E_M^{~A}$ and $Z^{\ul M} $.  An important property of the action (4.5) is
that it is invariant under $n=8$ local  world-line supersymmetry, as opposed to
the $\kappa$--symmetry. (The latter emerges as a special case of the former
in a certain gauge). The supersymmetry of the second and third terms in the
action is manifest (everything transform like supertensors), while the
supersymmetry of the first term is due to the fact that $E_r^{\ul a}$
transforms homogeneously like $D_r$ does, and this can be compensated by a
suitable transformation of the Lagrange multiplier.

The field equation for
$P_{\ul a}{}^r$ yields, as before, Eq. (3.13), and as an integrability
condition, the twistor constraint (3.14). The field equation for $P^M$ implies
that ${\tilde B}_M=\partial_M Q=0$, from which it follows that $d{\tilde
B}\equiv {\tilde H}$=0. Using the constraints (4.2), (4.4) and (3.13), one can
show that this constraint is indeed satisfied. In fact,
the form of ${\tilde B}$ is
engineered precisely such that $d{\tilde B}=0$,  modulo the constraints
(4.2),(4.4) and (3.13).

As a consequence of $d{\tilde B}=0$, the action (4.5) has also the gauge
invariance
$$
\delta P^M=\partial_N \Lambda^{NM}\ , \eqno(4.7)
$$
where $\Lambda^{MN}$ is an arbitrary graded antisymmetric superfield. In
showing this one uses the fact that $d{\tilde B}=0$, which in turn requires the
use of the constraint (3.13), which is the equation of motion for $P_{\ul
a}{}^r$. Of course, one is not supposed to use field equations in showing gauge
invariance. However, here we are allowed to do so, because in showing the
invariance, the terms that are proportional to the field equation of
$P_{\ul a}{}^r$ can always be cancelled by an appropriate variation of
$P_{\ul a}{}^r$.

The action (4.5) has the additional gauge invariance
$$
\delta P_{\ul a}^r = D_q\bigl(\xi^{qrs}\Gamma_{\ul a} E_s\big)\ ,\quad\quad
\delta P^M = -E_r{}^M D_q\bigl(\xi^{qrs} E_s\big)\ , \eqno(4.8)
$$
where the parameter $\xi^{qrs}_{\ul\alpha} (\tau,\theta)$ is totally
symmetric and traceless in its worldline indices, and we have used the
constraints (4.2), (4.4), (3.13) and assumed the existence of the Dirac matrix
identity
$$
\Gamma^{\ul a}_{\ul{\alpha\beta}}\Gamma^{\ul a}_{\ul{\gamma\delta}}
+C_{\ul{\alpha\beta}}C_{\ul{\gamma\delta}}
+{\rm cyclic}\ (\ul{\alpha\beta\gamma}) =0\ . \eqno(4.9)
$$
Among the spaces listed in the Table, this identity holds in $d=5,9$. The
gauge invariance (4.8) plays an important role in getting rid of many redundant
fields and thus in showing the classical equivalence of the action (4.5) with
the usual action (4.1).

Lets us now consider the remaining equations of motion. The field equation of
$Q$ reads $\partial_M P^M=0$. This equation has the solution [13]
$P^M=\partial_N\Sigma^{NM}+\theta^8\delta_\tau^M T$, where $T$ is a constant
and
$\Sigma^{MN}$ is an arbitrary graded antisymmetric superfield, which can be
gauged away by using the gauge symmetry (4.7). Substituting this algebraic
solution into the action (4.5) and after some algebra, one can show that the
action reduces to
$$
S=\int d\tau\bigg[ p_{\ul a}\big({\cal E}_0^{\ul a}
-\ft18 {\bar \lambda}_r\Gamma^{\ul a} \lambda_r \big)
 +\partial_\tau Z^{\ul M}B_{\ul M}
+\big({\cal E}_\tau^{\ul a}{\cal E}_\tau^{\ul a}\big)^{1/2}\bigg] \ ,
\eqno(4.10)
$$
where $p_{\ul a}=(D^7)_r P_{\ul a}^r\vert_{\theta=0}$. With arguments
parallel to those of [12,13], we expect that the  Lagrange multiplier
$p_{\ul a}$ does not describe any new degree of freedom, and the field
equations of (4.1) and (4.5) are classically equivalent.

The key ingredient in
the above formulation of massive superparticle was the existence a super
one-form ${\tilde B}$ on the worldline such that $d{\tilde B}=0$ modulo an
acceptable set of worldline and target superspace constraints. Therefore, in
order to generalize the above construction to higher super $p$--branes, it is
natural to search for closed $(p+1)$--forms in the worldvolume superspace
together with a suitable set of constraints in worldvolume and targe
superspaces. Indeed, in [16] we found such superforms and we were able to give
a general construction of the twistor-like super $p$--brane actions,
generalizing a result of [15] for the case of supermembrane. In the remainder
of this review, we shall summarize the result for general super $p$--branes.
\bigskip
\noindent{\bf 5.\ Twistor-like Formulation of Super $p$--Branes}
\bigskip
 In accordance with the procedure described earlier, we first fix the
supergeometry of the worldvolume superspace. In analogy with (3.1), we impose
the following constraints
$$
T_{\alpha\beta}{}^a=-2i (\Gamma^a)_{\alpha\beta},\qquad
T_{b\alpha}{}^a=0,\qquad T_{bc}{}^a=0,\qquad
T_{\alpha\beta}{}^{\gamma}=0\ . \eqno(5.1)
$$
See the Table for the symmetry properties of the gamma matrices. We also fix
the target superspace geometry. As for the target superspace geometry, in
addition to the superstorsion, we need to consider the Kalb-Ramond type super
$p+1$ form $B$ with field strength $H=dB$. In analogy with (4.2), we then
choose the following constraints [3]
$$
\eqalign{
&T_{\ul{\alpha\beta}}{}^{\ul c}= -2i(\Gamma^{\ul c})_{\ul{\alpha\beta}}\
,\quad\quad T_{\ul{b\alpha}}{}^{\ul a}=0\ , \quad\quad
T_{\ul{\alpha\beta}}{}^{\ul\gamma}=0\ , \cr
&H_{\ul{\alpha\beta c_1}...\ul{c_p}}=
i\xi^{-1}\big(\eta\Gamma_{\ul{c_1}...\ul{c_p}}\big)_{\ul{\alpha\beta}},\qquad
H_{\ul{\alpha b_1}...\ul{b_{p+1}}}=0,\qquad
H_{\ul{\alpha\beta\gamma}...\ul{A_1}...{\ul{A_{p-1}}}}=0\ , \cr} \eqno(5.2)
$$
where $\xi=(-)^{(p-2)(p-5)/4}$ and $\eta$ is a matrix chosen such that
$\eta\Gamma_{\ul{c_1}...\ul{c_p}}$ is symmetric. $\eta=1$ except for the
following cases: $\eta=\Gamma_{d+1}$ for $(p=3, d=8)$, with the definition
$\Gamma_{d+1}=\Gamma_0\Gamma_1\cdots \Gamma_{d-1}$, and $\eta=1\times
\sigma_2$ for $(p=2, d=5)$. See the Table for further information on the
notation and properties of the Dirac matrices in diverse dimensions.

In $d=11$ dimensions the above constraints
describe the $d=11$ supergravity theories.
In other cases, a detailed
analysis of the constraint remains to be carried out. Presumably, they
describe supergravity theories containing $(p+1)$--form potentials.

Having specified the geometry of the worldvolume and target superspaces,
our next goal is to write down an action for
twistor--like super $p$--branes in analogy with the action (4.5). Such an
action was proposed in [15] for the case of the supermembrane.
In [16]
we generalized that result and proposed the following action for all super
$p$--branes
$$
S= \int d^{p+1}\sigma d^{mn}\theta\bigg[ P_{\ul a}^\alpha
E_\alpha^{\ul a}  +P^{M_1\cdots M_{p+1}}\big({\tilde B}_{M_1\cdots M_{p+1}}
-\partial_{M_1} Q_{M_2\cdots M_{p+1}}\big)\bigg]\ ,  \eqno(5.3)
$$
where $P_{\ul a}^{\alpha r}$, $P^{M_1\cdots M_{p+1}}$ and
$Q_{M_1\cdots M_p}$ are Lagrange multiplier superfields (the latter two
are graded totally antisymmetric) and the $(p+1)$--form ${\tilde B}$ is
given by [16]
$$
\eqalign{
{\tilde B}_{M_1\cdots M_{p+1}}=&(-1)^{\epsilon_{p+1}(M,\ul{M})}\
 \partial_{M_{p+1}}Z^{\ul{M_{p+1}}}\cdots
 \partial_{M_1}Z^{\ul M_1} B_{\ul{M_1}\cdots \ul{M_{p+1}}}\cr
  &-{i\over 2mn(p+1)}\Gamma^{\alpha\beta}_{c_{p+1}}
          \bigg(E_{M_{p+1}}^{~c_{p+1}}\cdots E_{M_1}^{~c_1}
H_{\alpha\beta c_1\cdots c_p}+ {\rm cyclic}
\ [M_1\cdots M_{p+1}] \bigg)\ .\cr} \eqno(5.4)
$$
The grading factor is given by
$ \epsilon_{p+1}(M,\ul{M})= \sum_{n=1}^{p}(M_1+\cdots M_n)(M_{n+1}+
\ul{M_{n+1}})$, and the pullback of $H$ by
$$
H_{A_1\cdots A_{p+2}}=(-1)^{\epsilon_{p+2}(A,\ul{A})}
E_{A_{p+2}}^{~\ul{A_{p+2}}}\cdots E_{A_1}^{~\ul{A_1}}
H_{\ul{A_1}\cdots \ul{A_{p+2}}}\ . \eqno(5.5)
$$

The field equation for $P_{\ul a}^\alpha $ is
$$
    E^{\ul a}_\alpha =0\ . \eqno(5.6)
$$
The integrability condition for this equation yields the analog of the
twistor constraint (3.14) for super $p$--branes, and it takes the form
$$
(E_\alpha\Gamma^{\ul a} E_\beta )=\Gamma^a_{\alpha\beta} E_a^{\ul a}\ .
\eqno(5.7)
$$
The field equation for $P^{M_1\cdots M_{p+1}}$ is
${\tilde H}_{M_1 \cdots M_{p+2}}
=\partial_{M_1}{\tilde B}_{M_2\cdots M_{p+2}} +{\rm cyclic}\ [M_1\cdots
M_{p+2}]\ =0$. Given ${\tilde B}$ as in (5.4), it is nontrivial to show
that this equation holds. A tedious calculation, which can be found in [16]
and we will not reproduce here, shows that this closure property indeed holds
for the cases $(p,m,n) = (2,2,8), (5,4,2), (2,2,4),
(3,4,1), (2,2,2)$ and $(2,2,1)$ (See the Table). The $p=2$ cases were already
considered in [15]. In these calculations, the following Dirac matrix identity
plays an important role [3,4]
$$
\Gamma^{\ul c}_{(\ul{\alpha\beta}} \big(\eta \Gamma^{\ul{c c_1}\cdots
\ul{c_{p-1}}}\big)_{\ul{\gamma\delta})}=0\ . \eqno(5.8)
$$

Since the equation $d{\tilde B}=0$ holds, the analog of the gauge invariance
(4.7) exists also for super $p$--branes, and reads:
$\delta P^{M_1\cdots M_{p+1}}=\partial_N \Sigma^{NM_1\cdots M_{p+1}}$,
where the parameter is completely graded antisymmetric. Using this symmetry,
the field equation for $Q^{M_1...M_p}$:\ $
\partial_{M_1} P^{M_1\cdots M_{p+1}}=0$, can be put into the form
$P^{M_1\cdots M_{p+1}}=T\epsilon^{m_1\cdots m_{p+1}}
\delta_{m_1}^{M_1}\cdots \delta_{m_{p+1}}^{M_{p+1}} \theta^{mn}$, where $T$ is
constant. Substituting this into the action and after considerable amount of
algebra which has been described in [16], one finds the result
$$
\eqalign{
S=& \int d^{p+1}\sigma d^{mn}\theta P_{\ul a}^\alpha
E_\alpha^{\ul a}  +{(p+1)!\over 2}\int d^{p+1}\sigma
\big(-{\rm det}\ E_m^{~\ul a}
E_n^{~\ul a}\big)^{1/2}\vert_{\theta=0}\cr
 &+\int d^{p+1}\sigma \epsilon^{m_1\cdots m_{p+1}}
\partial_{m_{p+1}}Z^{\ul M_{p+1}}\cdots
 \partial_{m_1}Z^{\ul M_1} B_{\ul{M_1}\cdots
\ul{M_{p+1}}}\vert_{\theta=0}\cr}\ ,\eqno(5.9)
$$

Going back to the original form of the action, the field
equation for  $Z^{\ul M}$ derived from it, may seem to describe a large number
of degrees of freedom. However, one expects a number of gauge invariances,
similar to (4.8), which ought to play an important role in reducing drastically
the true number of degrees of freedom. In fact in [15], such
gauge invariances have been proposed for the case of supermembranes ($p=2$),
and it has been claimed that the true degrees of freedom are those that follow
from the usual $\kappa$--symmetric formulation of the supermembrane. We
have  not checked this, and we don't know yet what the full set of gauge
symmetries involving the Lagranage multiplier fields are, and consequently we
don't know yet what the true degrees of freedom are for general super
$p$--branes.
\bigskip
\noindent{\bf 6. Conclusions and Open Problems}
\bigskip
The main result concerning the twistor-like formulation of super $p$--branes
is the action (4.5), together with the definition (5.3), or alternatively
the formula (5.9). The latter form of the action coincides with the Nambu-Goto
form of the usual super $p$--brane action. The difference is due to the
Lagrange
multiplier term. It is not altogether clear whether the equations of motions
are equivalent to those which follow from the  usual super $p$--brane action.
For this to happen, one must show that there is sufficiently powerful gauge
symmetry of the action which makes it possible to gauge away the Lagrange
multiplier. We have shown that for the massive superparticle such a gauge
symmetry indeed exists. The existence of this gauge symmetry
relies on the Dirac matrix identity (4.9). It remains to be seen whether a
similar gauge symmetry exists for other values of $p$. We expect that the
$p$--brane Dirac matrix identity (5.8) will play an essential role in proving
the existence of such a symmetry.

One of the essential ingredients of the
twistor-like transform is the existence of  a closed super (p+1)-form on the
worldvolume superspace which is constructed out of the pull-backs of a super
$(p+1)$-form and its curvature in target superpspace. We have shown that this
closed $(p+1)$-form exists for the cases $(p,m,n) = (2,2,8), (5,4,2), (2,2,4),
(3,4,1), (2,2,2)$ and $(2,2,1)$. The $p=2$ cases were considered in
[15]. We believe that the existence of this closed
$(p+1)$-form should have some interesting geometric interpretation, independent
of the role it plays in the twistor-like transform. For instance, it seems
that it is related to the light-like integrability principle
[19,13].

There are a number of open problems which deserve further investigation. Some
of these problems are:

\item{(1)} What is the full set of symmetries of the action and what are
the  physical degrees of freedom?

\item{(2)} What is the precise relation between  our action and the usual one
[3] at the quantum level?

\item{(3)} Can the quantization problems of the usual $\kappa$--symmetric
action be avoided by the new action?

\item{(4)} Are the symmetries of the action anomaly-free?

\item{(5)} Is the twistor-like formulated of super $p$--brane theory finite?
Can one have a handle on this problem, at least at the perturbative level?

\item{(6)} How can we couple
Yang-Mills sector to super $p$--branes?  (Such theories are
usually referred to as heterotic $p$--brane theories, because of their
similarity to the heterotic string theory).

\np
\vskip 1.5truecm
\centerline{\bf REFERENCES}
\bigskip
\item{[1]} W. Siegel, Phys. Lett. {\bf B128} (1983) 397; Class. Quantum
          Grav. {\bf 2} (1985) 195;
\item{} M.B. Green and J.H. Schwarz, Phys. Lett. {\bf B136 } (1984) 367.
\item{[2]} J. Hughes, J. Liu and Polchinski, Phys. Lett. {\bf B180} (1986)
370.
\item{[3]} E. Bergshoeff, E. Sezgin and P.K. Townsend, Phys. Lett. {\bf B189}
(1987) 75.
\item{[4]} A.~Ach\'ucarro, J.M.~Evans, P.K.~Townsend and D.L.~Wiltshire, Phys.
Lett. {\bf B198} (1987) 441.
\item{[5]} M.J. Duff, Class. Quant. Grav. {\bf 5} (1988) 189;
\item{} A. Strominger, Nucl. Phys. {\bf B343} (1990) 167;
\item{} M.J. Duff and J.X. Lu, Nucl. Phys. {\bf B354} (1991) 141;
        Phys. Rev.	Lett. {\bf 66} (1991) 1402; Class. Quant. Grav.
       {\bf 9} (1991) 1;
\item{} C.G. Callan, J.A. Harvey and A. Strominger, Nucl. Phys.
      {\bf B359}(1991) 611; Nucl. Phys. {\bf B367} (1991) 60.
\item{[6]} L. Brink and J.H. Schwarz, Phys. Lett. {\bf B100} (1981) 310.
\item{[7]} A. Ferber, Nucl. Phys. {\bf 132} (1978) 55;
\item{} T. Shirafuji, Progr. Theor. Phys. {\bf 70} (1983) 18;
\item{} I. Bengston and M. Cederwall, Nucl. Phys. {\bf B302} (1988) 81.
\item{[8]} P.K. Townsend, Phys. Lett. {\bf 261} (1991) 65.
\item{[9]} N. Berkovits, Phys. Lett. {\bf B247} (1990) 45.
\item{[10]} D.P. Sorokin, V.I. Tkatch and D.V. Volkov, Mod. Phys. Lett. {\bf
A4}
         (1989) 901;
\item{} D.P. Sorokin, V.I. Tkacth, D.V. Volkov and A.A. Zheltukhin,
          Phys. Lett. {\bf B216} (1989) 302.
\item{[11]} F. Delduc and E. Sokatchev, Class. Quantum. Grav. {\bf 9} (1992)
         361;
\item{} P.S. Howe and P.K. Townsend, Phys. Lett. {\bf B259} (1991) 285;
\item{} A.S. Galperin, P.S. Howe and K.S. Stelle, Nucl. Phys. {\bf B368}
          (1992) 281;
\item{} P.S. Howe and P.C. West, Int. J. Mod. Phys. {\bf A7} (1992) 6639.
\item{[12]} A. Galperin and E. Sokatchev, Phys. Rev. {\bf D46} (1992) 714.
\item{[13]} E. Delduc, A. Galperin, P.S. Howe and E. Sokatchev, Phys. Rev.
          {\bf D47} (1993) 578.
\item{[14]} N. Berkovits, Phys. Lett. {\bf B232} (1989) 184;
\item{} F. Delduc, E. Ivanov and E. Sokatchev, Nucl. Phys. {\bf B384} (1992)
        334;
\item{} M. Tonin, Phys. Lett. {\bf B266} (1991) 312; Int. J. Mod. Phys.
            {\bf A7} (1992) 6013;
\item{} N. Berkovits, Nucl. Phys. {\bf B379} (1992) 96;
\item{} D.P. Sorokin and M. Tonin,
          preprint, DFPD/93/TH/52 (hep-th/9307039);
\item{} A. Galperin and E. Sokatchev,
          preprint, BONN-HE-93-05 (hep-th/9304046);
\item{} P. Pasti and M. Tonin, preprint (hep-th/9405074).
\item{[15]} P. Pasti and M. Tonin, preprint, DFPD/93/TH/07 (hep-th/9303156).
\item{[16]} E. Bergshoeff and E. Sezgin, Nucl. Phys. {\bf B422} (1994) 329.
\item{[17]} I.A. Bandos and A.A. Zheltukhin, Int. J. Mod. Phys. {\bf A8}
(1993) 1081.

\item{[18]} E. Sezgin, preprint, CTP TAMU-28/93 (hep-th/9310126).
\item{[19]} E. Witten, Nucl. Phys. {\bf B266} (1986) 241.


\np

\centerline{TABLE}

\bigskip

 $$
\vbox{\def\tablerule{\noalign{\hrule}}
\offinterlineskip\baselineskip=16pt\halign{\strut#&
\vrule#&
\hfil#\hfil&\vrule#&\hfil#\hfil&\vrule#&\hfil#\hfil&\vrule#&
\hfil#\hfil&\vrule#&\hfil#\hfil&\vrule#&\hfil#\hfil&\vrule#&
\hfil#\hfil&\vrule#&\hfil#\hfil&\vrule#&\hfil#\hfil&\vrule#&
\hfil#\hfil&\vrule#&\vrule#\cr
\tablerule
     &&&\omit&&\omit&&\omit&&\omit&&\omit&&\omit&&\omit
     &&\omit&&\cr
&&\multispan{19}\hfil Target Space Data \hfil &\cr
     &&&\omit&&\omit&&\omit&&\omit&&\omit&&\omit&&\omit&&\omit
     &&\omit&&\cr
\tablerule
&&$d$&&$11$&&$10$&&$9$&&$8$&&$7$&&$6$&&$5$&&$4$&\cr
\tablerule
&&\ $(M,N)\ $&&(32,1)&&(16,1)&&(16,1)&&(16,1)&&(8,2)
&&(4,2)&&(4,2)&&(4,1)&\cr
\tablerule
&&${\ul G}$&&--&&--&&--&&--&&\ USp(2)\ &&\ USp(4)\ &&\ USp(4)\
&&\ SO(4)\ &\cr
\tablerule
&&$C_{\ul{\alpha'\beta'}}$&&A&&A&&S&&S&&S&&S&&S&&A&\cr
\tablerule
&&$\Gamma^{\ul a}_{\ul{\alpha'\beta'}}$&&S&&S&&S&&S &&A&&A
&&A&&S&\cr
\tablerule
&&$\eta_{\ul{rs}}$&&--&&--&&--&&--&&A&&A&&A&&S&\cr
\tablerule
&&Type&&M&&MW&&PM&&PM&&SM&&SMW&&SM&&M&\cr
\tablerule
 &&&\omit&&\omit&&\omit&&\omit&&\omit&&\omit&&\omit
    &&\omit&&\omit&&\cr
&&\multispan{19}\hfil Worldvolume  Data ($p\ge 2$) \hfil&\cr
    &&&\omit&&\omit&&\omit&&\omit&&\omit&&\omit&&\omit
    &&\omit&&\omit&&\cr
\tablerule
&&$p$&&$2$&&$5$&&$4$&&$3$&&$2$&&$3$&&$2$&&$2$&\cr
\tablerule
&&$(m,n)$&&(2,8)&&(4,2)&&(4,2)&&(4,2)&&(2,4)&&(4,1)
&&(2,2)&&(2,1)&\cr
\tablerule
&&$G$&&\ SO(8)\ &&\ USp(2)\ &&\ USp(2)\
&&\ SO(2)\ &&SO(4)&&--&&SO(2)&&-- &\cr
\tablerule
&&$C_{\alpha'\beta'}$&&A&&S&&S&&A&&A&&A&&A&&A&\cr
\tablerule
&&$\Gamma^a_{\alpha'\beta'}$&&S&&A&&A&&S&&S&&S&&S&&S&\cr
\tablerule
&&$\eta_{rs}$ &&S&&A&&A&&S&&S&&--&&S&&--&\cr
\tablerule
&&Type&&M&&SMW&&SM&&M&&M&&M&&M&&M&\cr
\tablerule}}
$$

\vskip 1.5truecm
In this table, $d$ indicates the
dimension of spacetime, $M$ is the dimension of the spinor irrep of
$SO(d-1,1)$, $N$ is the dimension of the defining
representation of the automorphism group ${\ul G}$ of the super Poincar\'e
algebra in $d$ dimensions, $C_{\ul{\alpha'\beta'}}$ is the charge
conjugation matrix, $\Gamma^{\ul a}_{\ul{\alpha'\beta'}}$ are the Dirac
matrices $(\Gamma^{\ul a}C)_{\ul{\alpha'\beta'}}$ and $\eta_{\ul{rs}}$ is the
invariant tensor of ${\ul G}$. We often use the notation in which a pair of
indices $({\ul \alpha'}r)$ is replaced by a single index ${\ul \alpha}$.
Furthermore, in $d=6,10$ the matrices $\Gamma^{\ul a}_{\ul {\alpha\beta}}$ are
chirally projected  Dirac matrices and $\Gamma_{\ul a}^{\ul{\alpha\beta}}$
are projected with opposite chirality. In this notation
raising or lowering of the spinor indices is not needed. The types of spinors
are characterized according to the reality and chirality conditions imposed on
them, namely Majorana (M), pseudo-Majorana (PM), symplectic Majorana (SM),
Majorana-Weyl (MW) and symplectic Majorana-Weyl (SMW). Corresponding
quantities are listed for the super $p$--branes that arise in target space
dimension $d$.

\end